\documentclass[natbib,preprint]{sigplanconf}

\newcommand*{\ARXIV}{}

\usepackage[hidelinks]{hyperref}

\usepackage{natbib}
\usepackage[table]{xcolor}
\usepackage{graphicx}
\usepackage{amsmath}
\usepackage{breqn}
\usepackage{accents}
\usepackage{subcaption}
\usepackage{xfrac}
\usepackage{xspace}
\usepackage{listings}
\usepackage{booktabs}
\usepackage{multirow}
\usepackage{amssymb}
\usepackage{todonotes}
\usepackage{ltxcmds}

\unless\ifdefined\ARXIV
    \graphicspath{{./images/}}
\fi

\begin{document}

\title{A high-level model of embedded flash energy consumption}
\authorinfo{James Pallister}{University of Bristol}{james.pallister@bristol.ac.uk}
\authorinfo{Kerstin Eder}{University of Bristol}{kerstin.eder@bristol.ac.uk}
\authorinfo{Simon J. Hollis}{University of Bristol}{simon.hollis@bristol.ac.uk}
\authorinfo{Jeremy Bennett}{Embecosm}{jeremy.bennett@embecosm.com}

\conferenceinfo{CONF 'yy}{Month d--d, 20yy, City, ST, Country}
\copyrightyear{2014}
\copyrightdata{978-1-nnnn-nnnn-n/yy/mm}
\doi{nnnnnnn.nnnnnnn}

\permissiontopublish             

\maketitle

\newcommand{\STMZERO}{STM32F0\xspace}
\newcommand{\STMTHREE}{STM32F1\xspace}
\newcommand{\ATMEGA}{ATMEGA328P\xspace}
\newcommand{\PIC}{PIC32MX250F128B\xspace}
\newcommand{\MSP}{MSP430F5529\xspace}
\newcommand{\FRAM}{MSP430FR5739\xspace}

\newcommand{\romann}[1]{{\sc\bf#1}}

\newcommand{\ubari}[0]{\hat\imath}
\newcommand{\ubarj}[0]{\hat\jmath}

\begin{abstract}

The alignment of code in the flash memory of deeply embedded SoCs can have a large impact on the total energy consumption of a computation. We investigate the effect of code alignment in six SoCs and find that a large proportion of this energy (up to 15\% of total SoC energy consumption) can be saved by changes to the alignment.

A flexible model is created to predict the read-access energy consumption of flash memory on deeply embedded SoCs, where code is executed in place. This model uses the instruction level memory accesses performed by the processor to calculate the flash energy consumption of a sequence of instructions. We derive the model parameters for five SoCs and validate them. The error is as low as 5\%, with a 11\% average normalized RMS deviation overall.

The scope for using this model to optimize code alignment is explored across a range of benchmarks and SoCs. Analysis shows that over 30\% of loops can be better aligned. This can significantly reduce energy while increasing code size by less than 4\%. We conclude that this effect has potential as an effective optimization, saving significant energy in deeply embedded SoCs.

\end{abstract}

\section{Introduction}

The demand of longer battery life, with increased functionality in our embedded systems motivates the need to improve the energy consumption of these devices.
This is particularly noticeable in \textit{deeply} embedded devices, whose battery we expect to last on the time scale of years. While previous attempts at reducing energy consumption focused on improving the hardware to prolong battery life, a software-centric approach is necessary to achieve maximal energy savings.

In these deeply embedded devices, there is typically a System on Chip (SoC) at the heart of the device, controlling the system. These SoCs are small devices without caches that often execute directly out of embedded flash memory. With current technologies allowing up to 8MB of embedded flash~\cite{RenesasElectronics2014}, the majority of silicon area and therefore a large proportion of the power dissipation is taken by the embedded memory.

Flash does not have a uniform structure, causing address dependent energy consumption. This paper looks at how this energy consumption can be modeled and then reduced, with minimal overhead in execution time and code size. This can be done by considering the code's absolute address in the flash memory and adjusting the code's position.

\begin{figure}[b!]
    \includegraphics[width=\linewidth]{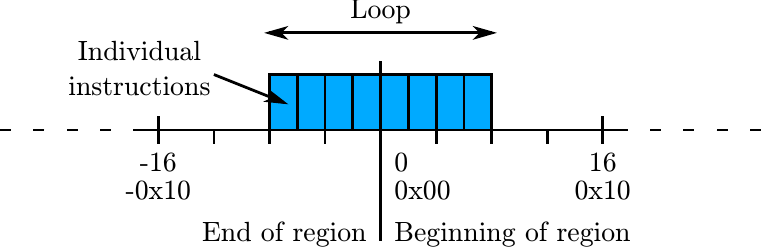}
    \caption{Illustration of a misaligned loop, causing additional energy consumption}
    \label{fig:introduction_example}
\end{figure}

An example of the way the absolute address of code in flash affects energy can be seen in Figure~\ref{fig:introduction_example}. This diagram shows a sequence of instructions, crossing a page boundary in flash. The crossing of this boundary causes additional energy consumption, due to additional circuitry being powered up to access the new page. If the code did not cross this boundary, the energy consumption of the code sequence would be lower, since there is no need to power up the support circuitry.

\begin{table*}
    \centering
    \begin{tabular}{p{3cm} l@{~}l r@{}p{2.05cm} r@{}p{1.7cm} r c}
        \toprule
        \bf\centering SoC   & \multicolumn{2}{c}{\bf Architecture}
        & \multicolumn{2}{p{2.2cm}}{\centering\bf Non-volatile memory}
        & \multicolumn{2}{p{2cm}}{\centering\bf Volatile memory}
        & \multicolumn{1}{p{0.7cm}}{\centering\bf Bit width}
        & \multicolumn{1}{p{1.7cm}}{\centering\bf Instruction bit width}\\
        \midrule
        STM32F051R8     & ARM & Cortex-M0 & 64&kB Flash          & 8&kB SRAM    & 32 & 16\hspace{.3mm}\footnotemark[1]\\
        STM32F100RB     & ARM & Cortex-M3 & 64&kB Flash          & 8&kB SRAM    & 32 & 16/32\hspace{.3mm}\footnotemark[2]\\
        ATMEGA328P      & Atmel & AVR    & 32&kB Flash          & 2&kB SRAM    & 8  & 16 \\
        PIC32MX250F128B & MIPS & M4K      & 128&kB Flash         & 32&kB SRAM   & 32 & 16/32\hspace{.3mm}\footnotemark[3] \\
        MSP430F5529     & TI & MSP430     & 128&kB Flash         & 32&kB SRAM   & 16 &  16     \\
        MSP430FR5739    & TI & MSP430     & 16&kB FRAM           &  1&kB SRAM   & 16 &  16     \\
        \bottomrule
    \end{tabular}
    \caption{Features of the SoCs selected.}
    \label{tab:socs}
    \scriptsize
    \raggedleft\footnotemark[1] With some 32 bit instructions. \\
    \raggedleft\footnotemark[2] ISA supports interleaved 16 and 32 bit instructions. \\
    \raggedleft\footnotemark[3] MIPS16e mode can be entered for 16 bit instructions. \\
\end{table*}

Compilers are an obvious target for this approach, being able to automatically apply code transformations for the developer. Implementing the transformation in a compiler also has the added benefit of energy efficiency upgrades with few modifications to the developer's source code; for the user, a trivial compiler version upgrade could implement new energy efficient optimizations.

Previous optimizations have considered how memory alignment affects energy consumption in caches, for both code and data~\cite{Calder1998,Chilimbi1999,Chandra2008}. Typically, ensuring a frequently executed piece of code is in a single cache line will reduce the energy consumption of the cache, since fewer cache lines are powered up and fewer cache misses occur~\cite{Kandemir2002a}. However, in deeply embedded systems, caches very rarely exist, due to power, size and cost constraints. While the principle of moving pieces of code to a better location is similar, the different structure of flash and its different energy consumption characteristics mean that the same techniques cannot be applied.

Another difference between embedded flash memory and caches is the diversity in embedded flash. The majority of caches operate similarly, and can be modeled simply by considering the cache-line size. In embedded flash memory there are a number of parameters which may have an effect on energy consumption, and these parameters vary from SoC to SoC. Furthermore, the characteristics depend on the SoC vendor's choice of flash technology --- similar processor architectures could be on die with very different flash architectures. This requires a generic model of flash memory to be constructed, with parameters that can be tuned to a wide range of embedded flash types. This paper considers six different SoCs and finds different energy consumption characteristics, even between similar processors.

The following contributions are made:

\begin{itemize}
    \item A model for flash memory energy consumption. This model considers read accesses only, as the code is infrequently modified these deeply embedded processors and a read-only model is sufficient to enable optimization. The model is applicable across a wide range of SoCs, and the parameters for each of these SoCs are found and explained with reference to the underlying structure of the flash memory.
    \item An analysis of how loop alignment in flash memory affects the energy consumption, including how the various features of the embedded flash correspond to the given model. The model is validated and shown to predict the energy consumption due to flash memory.
    \item An analysis of the scope for energy optimizations in deeply embedded processors using this model. A transformation is justified by statically analyzing a benchmark suite, showing that 30--40\% of all loops would benefit from the optimization, with less than 4\% increase in code size.
\end{itemize}

This paper is structured as follows. The following section gives a description of the SoCs used, and the structure of flash memory. Section~\ref{sec:modelling} presents the model for energy consumption of flash memory and Section~\ref{sec:loop_alignment} discusses tests and measurement collection from the previously listed SoCs. Following this, in Section~\ref{sec:regression} the model parameters for each SoC are derived and discussed. Then, a possible optimization and its justification with an analysis of a benchmark suite is given in Section~\ref{sec:optimization_scope}. Section~\ref{sec:related_work} discusses related work in this area, and, finally, Section~\ref{sec:conclusion} presents the conclusion to this paper.

\section{Platforms}

The proposed techniques are evaluated on several different SoCs, to demonstrate their portability. These platforms cover a range of deeply embedded processors, with a variety of instruction sets and SoC configurations.

It is necessary to distinguish between the instruction set, the architecture \textit{and} the hardware implementation in a SoC for the purposes of this paper. The architecture and instruction set are not enough to identify the energy consumption characteristics that occur because embedded flash with different structures may be included with the same processor. This results in characteristics which are specific to the combination of architecture and flash structure.

The chosen processors have a variety of different instruction widths, and some are variable length. This covers a spread of different types of instruction set, and causes various different code alignments when compiling for each architecture. It is hypothesized that this will provide good coverage of energy consumption effects due to flash memory and expose any alignment affects that may occur on these platforms.

The SoCs used in this paper are described below, and important features are shown in Table~\ref{tab:socs}.

\begin{description}
    \item[\STMZERO] ARM Cortex-M0. This SoC has a popular 32 bit processor that mostly executes 16 bit instructions.
    \item[\STMTHREE] ARM Cortex-M3. The Cortex-M3 processor is similar to the Cortex-M0 but executes a superset of instructions, including more 32-bit instructions.
    \item[\ATMEGA] Atmel AVR. This is an 8-bit processor, with instructions which are 16-bits long.
    \item[\PIC] Microchip PIC (MIPS). This processor was chosen for its use of the MIPS M4K core, and its direct access to the flash with no cache. This core also supports the 16-bit MIPS16e instruction set.
    \item[\MSP] TI MSP430. This is a 16-bit DSP processor, with a 16-bit instruction set. However some instructions can be up to $3\times 16$-bits long.
    \item[\FRAM] TI MSP430. This device has an identical processor architecture to the above SoC, and minor modifications to the peripherals and memory sizes. However, the defining feature is it uses FRAM instead of flash as its non-volatile storage. Direct comparison with the previous processor should allow effects due to difference in memory to be exposed.
\end{description}

The aim behind using this mix of SoCs is to demonstrate the differences in embedded flash, and the confounding effects that the processor architecture has on energy consumption.

\section{Flash Memory Structure}

\begin{figure}
    \centering
    \includegraphics[width=0.7\linewidth]{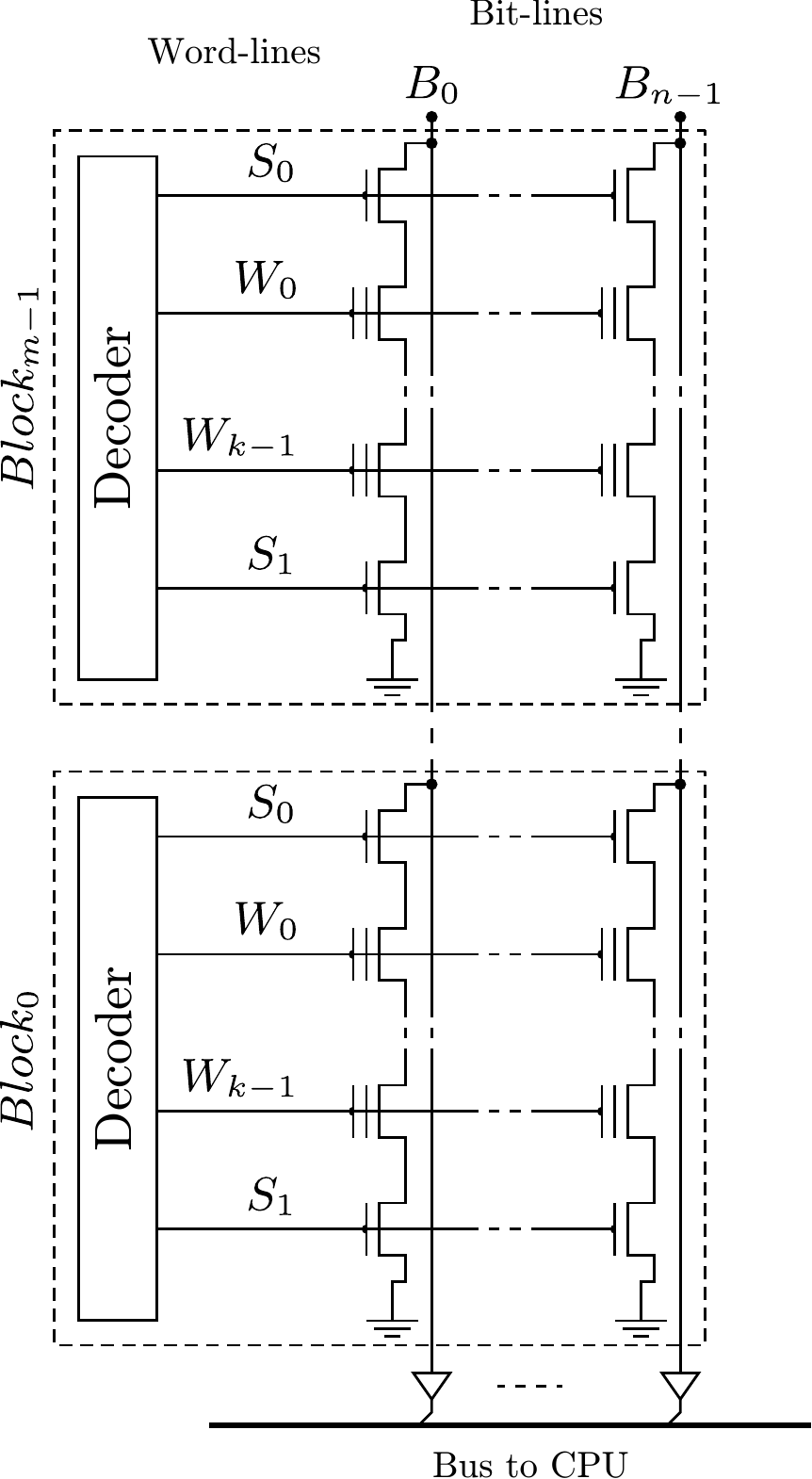}
    \caption{Structure of a single page in embedded flash memory.}
    \label{fig:structure_embedded_flash}
\end{figure}

The majority of embedded flash used in modern, deeply embedded SoCs is embedded NAND flash, chosen for its high density. The typical disadvantages associated with NAND flash over NOR flash are that NAND flash can only be erased and programmed in large blocks, while NOR flash can be programmed at a much finer granularity~\cite{Abari2008}. However, this disadvantage is generally insignificant for deeply embedded applications, due to the infrequent need to update the firmware.

An example of embedded flash structure is shown in Figure~\ref{fig:structure_embedded_flash}. This diagram shows a page of flash, containing individual flash cells arranged into word-lines and bit-lines~\cite{Cavaleri2004}, then formed into blocks~\cite{STMicroelectronics2005}. The bit-line size is typically 16 or 32 bits ($n$ in the diagram), and word line is typically 4 or 8 words ($k$ in the diagram) per block~\cite{Abari2008} (with $m$ blocks per page).

This structure allows entire word lines to be read simultaneously by selecting the block, and particular word line. The bit-lines are then charged and the select gates ($S_0$ and $S_1$) are used to connect the block to the bit-lines. Each sense amplifier on the bit-lines is used to read the flash cell's value, and propagate the bit's value onto the SoC's interconnect.

One key consequence of accessing the flash array $n$ bits at a time is that unaligned accesses must perform two reads each, powering up different word-lines and extract the relevant parts to return to the processor. This will result in additional energy consumption, and higher power dissipation if both reads must be performed in the same cycle.

When changing from one page to another, there will be a large associated energy cost, as additional sense amplifiers and decoder circuitry will be powered up. If code is executed directly from flash, when execution changes from one flash page to another, a measurable increase in the total energy consumption should occur.

It is hypothesized that the layout of flash memory will have a significant effect on the energy consumption of code executing out of it. The specifications of the embedded flash in modern SoCs are not generally available, thus it is hard to create an analytical model of the flash. In Section~\ref{sec:modelling}, a model is created with parameters that can be empirically tuned to a specific SoC with embedded flash.

\subsection*{FRAM}
\label{sec:fram}

Ferroelectric RAM is a newer technology that has lower energy consumption and different access characteristics compared to flash~\cite{Kato2010}. In particular, the structure of FRAM is different and can be accessed in a truly random fashion, as opposed to the word-lines and blocks of flash memory~\cite{Abari2008}. It is expected that the alignment of the executing code will have less effect on the energy consumption of the FRAM SoC, in comparison to the flash SoCs.

The \FRAM SoC (see Table~\ref{tab:socs}) was chosen because it uses this type of memory instead of flash.

\section{Modeling}
\label{sec:modelling}
\begin{figure}
    \includegraphics[width=\linewidth]{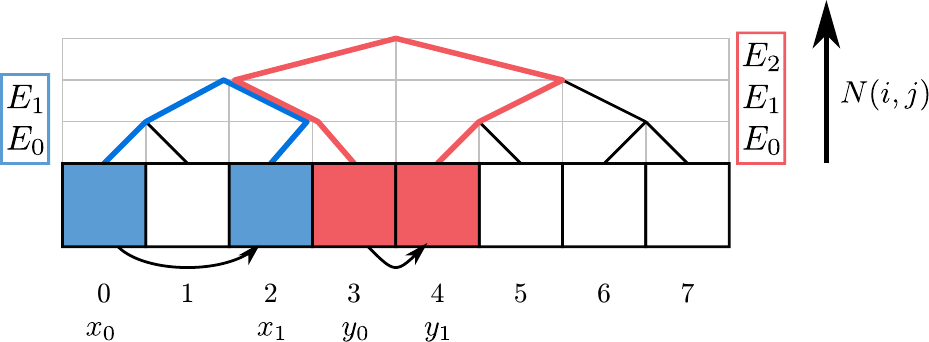}
    \caption{The regions changed when $A(x_0)\rightarrow A(x_1)$ and $A(y_0)\rightarrow A(y_1)$. The definition of $N(i,j)$ is given in Eq.~\ref{eq:Nij}.}
    \label{fig:region}
\end{figure}

This section discusses how the energy consumption caused by the flash structure can be modeled. Due to the prevalence of execute-in-place for deeply embedded microcontrollers, only read access for code execution is considered.

It is hypothesized that each time a consecutive flash memory access changes between arbitrary $2^k$-byte regions, there will be an associated energy cost $E_k$. The cost is cumulative: if a 4-byte region is changed, then a 2-byte and a 1-byte region will also have been changed. This forms a generic model that can be applied to a variety of different processors with embedded flash. For example, this could model the powering up of a different decoder every 16 bytes, along with an energy cost every 128 bytes for changing pages. These memory accesses are directly related to the instructions executing out of flash memory. Due to the undocumented sizes of various flash array structures, such as the number of bit-lines and word-lines, the model must be kept generic to ensure its applicability.

The following examples (shown in Figure~\ref{fig:region}) illustrate how the transition between two memory locations will utilize different model parameters (given by $E_0,E_1,...,E_k$). The full model is given in Eq.~\ref{eq:generic_flash_line_model}. For example, if an access $x_0$ is at address $x_0=0$ and the next access is at $x_1=2$, then both a one-byte boundary and a two-byte boundary have been crossed. Therefore the energy cost for this transition is represented by:
\begin{equation}
    x_0\rightarrow x_1 = E_0+E_1,
\end{equation}

where $E_0$ is the energy cost for crossing a one-byte boundary and $E_1$ is the energy cost for crossing a two-byte boundary.

Similarly, if $y_0=3$ and $y_1=4$, the energy cost will be:

\begin{equation}
    y_0\rightarrow y_1 = E_0+E_1+E_2.
\end{equation}

This can be abstracted to arbitrary accesses $i$ and $j$ in the following equation:

\begin{equation}
    i\rightarrow j= \sum_{k=0}^{N(i,j)}E_k,
\end{equation}

where $i$ and $j$ represent memory addresses. The term $N(i,j)$ represents the largest region that has been changed ($2^{N(i,j)}$ bytes), and therefore all smaller regions must have also changed. This is given by:

\begin{equation}
    N(i,j) = \left\lfloor\textrm{log}_2\Big(i\oplus j\Big)\right\rfloor.
    \label{eq:Nij}
\end{equation}

The symbol $\oplus$ is the bitwise exclusive-or between the two addresses.

The expression for a single transition can be built up into the entire memory energy cost for an application, $T$, by considering all accesses to the flash,

\begin{equation}
    E(T) = \sum_{(M_i,M_j) \in M}\left(M_i\rightarrow M_j\right),
    \label{eq:generic_flash_line_model}
\end{equation}

where $M_i$ and $M_j$ are consecutive flash accesses, and $M$ is the set of all consecutive accesses. In this form, the model requires detailed information about every memory read, whether from instruction fetch or data access. It can be challenging to analyze data accesses statically, therefore an approximation to the model can be made by noting:


\begin{equation}
    C \subset M,
\end{equation}

where $C$ is the set of accesses performed by sequentially executed instructions. This forms the following approximation to the model:

\begin{equation}
    E(T) \approx \sum_{(\ubari,\ubarj) \in C}\left(A(\ubari)\rightarrow A(\ubarj)\right),
    \label{eq:flash_line_model}
\end{equation}

where $\ubari$ and $\ubarj$ are sequential instructions and $A(\ubari)$ determines the address of instruction $\ubari$.

By ignoring the data accesses to the flash, the accuracy of the instruction-access model will be lower than the model in~\ref{eq:generic_flash_line_model}, however, it enables easy analysis of program at compile time.

The parameters $E_k$ can be characterized by measuring the energy consumption of carefully placed instructions in flash or by other methods, such as linear regression. Once the parameters have been found, the cost of moving code to different addresses in memory can be explored. The model can potentially be used as a heuristic in compiler optimizations, allowing strategic code placement to reduce the flash-access energy cost.

\subsection*{Instruction Fetching}
\label{sec:fetching}

\begin{figure}
    \centering
    \includegraphics[width=0.7\linewidth]{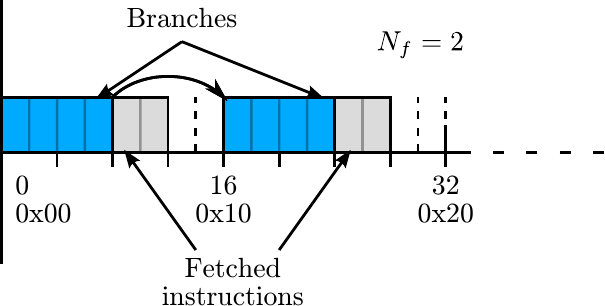}
    \caption{Diagram of memory addresses accessed due to instruction fetching.}
    \label{fig:insn_fetch}
\end{figure}

This section discusses how the instruction fetching performed by the cores will affect the energy consumption, and how the instruction-access model can be improved to account for this.

All of the processors tested are pipelined --- while they are executing the current instruction, they are fetching at least the next one to be executed. Since this happens for every instruction, the additional memory accesses do not affect the sequence of memory accesses, except when a branch is taken. A taken branch will have caused additional memory accesses, which will not have been taken into account from the model in Eq.~\ref{eq:flash_line_model}. Fig.~\ref{fig:insn_fetch} shows the additional instructions fetched.

The instruction-level model can be modified to account for these additional memory accesses. The following expression for $P$ describes the memory accesses due to fetching.

\begin{equation}
    P = \bigcup_{\forall (\ubari,\ubarj)\in C_b}\Big\{(\ubari+k,\ubari+k+1)\hspace{0.5mm}\Big|\hspace{0.5mm}k=0,...,N_f-1\Big\},
\end{equation}

where $(\ubari,\ubarj)$ is a taken branch instruction from instruction $\ubari$ to $\ubarj$ from the set of branches, $C_b$, and $N_{f}$ is the number of additional instructions fetched by the processor. This formula describes the $N_{f}$ extra instruction transitions needed for branch instruction $\ubari$ (which independent of the branch destination, $\ubarj$). The energy consumption due to instruction fetch ($E_f(T)$) can then be calculated with the following formula:

\begin{equation}
    E_f(T) = \sum_{(\ubari,\ubarj) \in \left( C \cup P\right)}\Big(A(\ubari)\rightarrow A(\ubarj)\Big),
    \label{eq:fetch_model}
\end{equation}

where $C$ is the set of consecutive instruction accesses (as used in Eq.~\ref{eq:flash_line_model}).

\begin{figure}[b!]
    \begin{subfigure}[t]{\linewidth}
        \includegraphics[width=\linewidth]{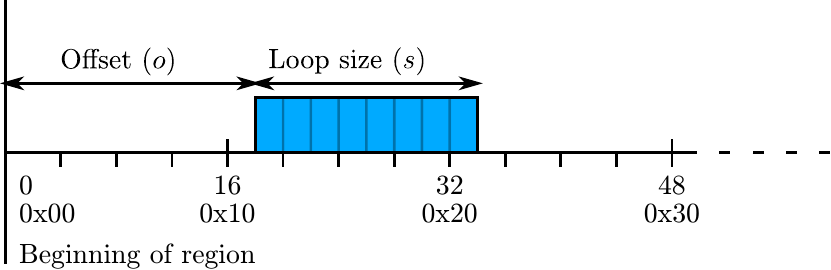}
        \caption{The offset and size of the loop relative to the beginning of the flash page.}
        \label{fig:loop_alignment_diagram}
    \end{subfigure}
    \vspace{5mm}

    \begin{subfigure}{\linewidth}
        \includegraphics[width=\linewidth]{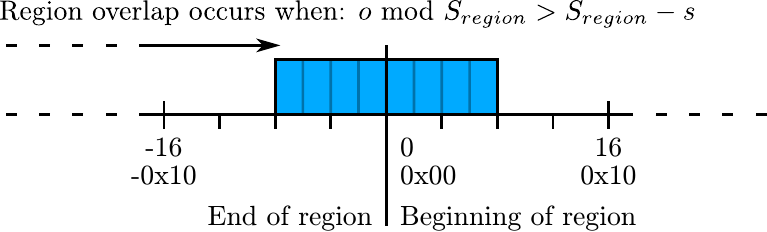}
        \caption{The loop crossing two flash regions.}
        \label{fig:loop_overlap}
    \end{subfigure}

    \caption{Diagrams of how loop alignment can be tested.}
        \label{fig:loop_alignment_diagrams}
\end{figure}

The amount of fetching performed by the processors is typically 1 or 2 instructions, and is listed in the relevant datasheet. When the extra terms incorporated into the model, it better fits the instruction sequences with loops in them (see Sections~\ref{sec:loop_alignment} and~\ref{sec:validation}).

The model's dependence on branching behavior means that the energy consumption is unknown until the conditional branches destinations are known. This causes the energy prediction to have an upper and lower bound, if analyzing the code statically. While this could decrease the accuracy of static predictions, in Section~\ref{sec:validation} the model is validated to make accurate predictions with conditional branching.


\section{Loop Alignment Tests}
\label{sec:loop_alignment}

This section discusses how the alignment of a loop affects the energy consumption of the SoC and describes tests performed to highlight the change in energy consumption. These tests are used in this section to demonstrate the effects of loop alignment and in the following section to derive the parameters in the model. The results in this section are actual measurements from the instrumented versions of each piece of hardware.

From the structure of flash, it is expected that the SoC's energy consumption will differ when code is executed from different addresses in its memory space. This was tested by choosing simple loops of different size and alignment (with respect to the beginning of memory), and measuring their energy consumption, as seen in Fig.~\ref{fig:loop_alignment_diagrams}. In this diagram, $S_{region}$ is the size of a page in flash, $o$ is the offset of the loop in memory and $s$ is the size of the loop, both in bytes. All are multiples of $2$ bytes, using 16-bit instructions for all platforms.

\begin{figure}[t]
    \includegraphics[width=\linewidth]{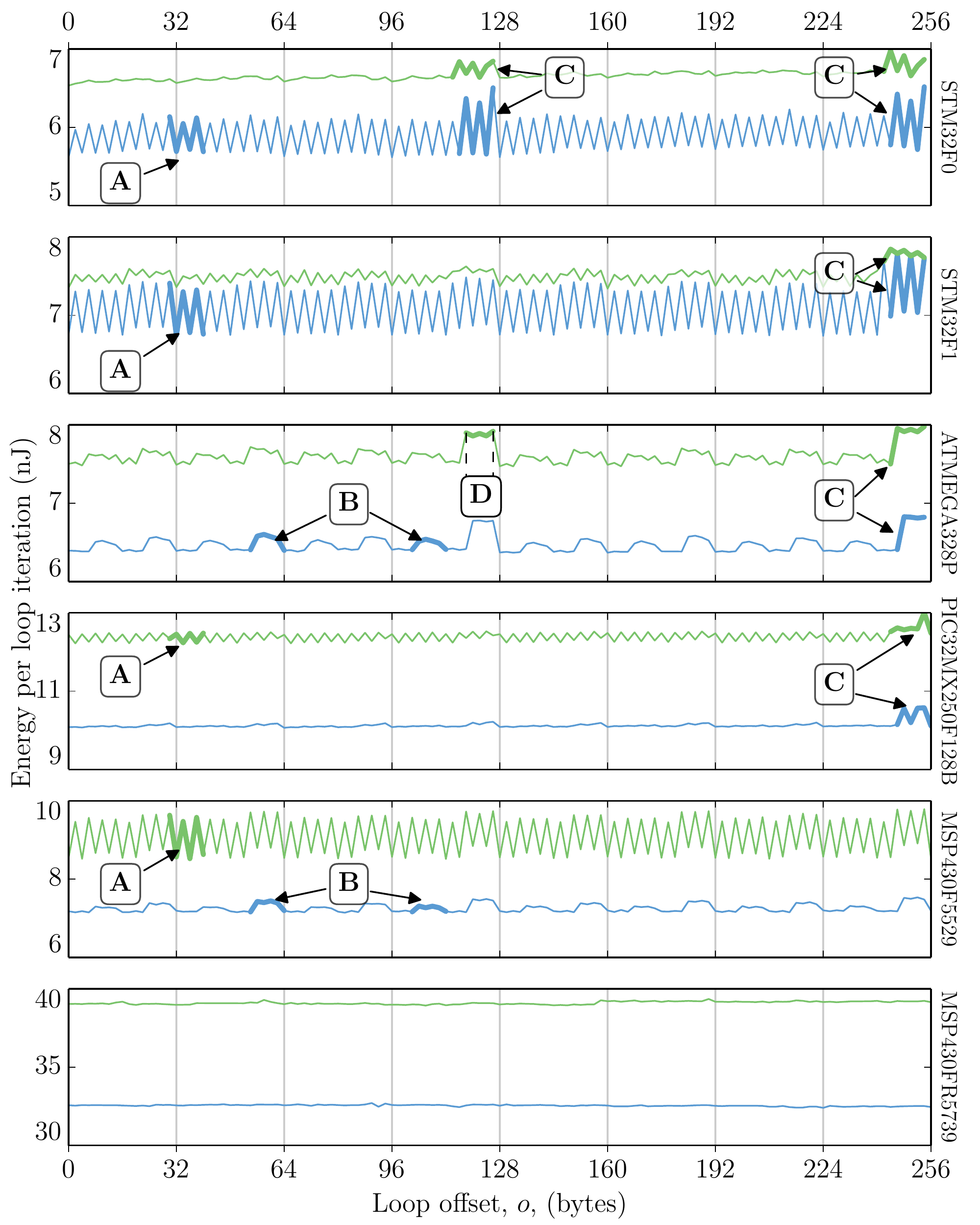}
    \caption{The effect of loop alignment on energy consumption for $S_{loop}=\{8,10\}$. In all cases $s=8$ is the lower line (blue) and $s=10$ is the upper line (green). The letters \textbf{A}--\textbf{D} represent features which can be directly mapped to parameters of the model. These are described more in the text and feature \textbf{D} is shown in more detail in Fig.~\ref{fig:feature_d}.}
    \label{fig:alignment_results}
\end{figure}


In the tests run, $T_{(o,s)} \in T$, where $o \in O_{loop}$ and $s\in S_{loop}$:
\begin{equation}
    O_{loop} = \left\{0,2,4,..,256\right\},
\end{equation}
\begin{equation}
    S_{loop} = \left\{8,10,12,...\right\}.
\end{equation}

Therefore, the set of tests covered is given by:
\begin{equation}
    T = O_{loop} \times S_{loop}.
\end{equation}

$T$ forms an exhaustive set of loop alignments and sizes, exposing alignment effects and providing sound data to derive the model parameters in Section~\ref{sec:regression}. The energy due to flash memory for each test, $E_f\left(T_{(o,s)}\right)$, can then be calculated from the model given earlier, in Eq.~\ref{eq:fetch_model}.



The actual energy consumption for each $T_{(o,s)}$ can be seen in Fig.~\ref{fig:alignment_results}, showing these tests repeated on multiple platforms for $S_{loop}=\{8,10\}$. The effects seen in this graph can be divided into four observations, which are combined differently in each of the graphs.

\begin{table*}
\centering
\begin{tabular}{l r r r r r r r r c}
    \toprule
    \multirow{2}{*}{SoC}         & \multicolumn{7}{c}{Model parameters (pJ)} \\
                & $E_2$ (\textbf{A}) & $E_3$ & $E_4$ (\textbf{B}) & $E_5$ & $E_6$ & $E_7$ (\textbf{C}) & $E_8$ (\textbf{C}) & $N_f$\\
    \midrule
    \STMZERO   & \cellcolor{red!80} 300 &  27 &  6 &  0 &  9 &  \cellcolor{red!40} 100 &  6  & 2\\
    \STMTHREE  & \cellcolor{red!80} 500 &  0 &  6 &  34 &  4 &  10 &  \cellcolor{red!40} 190  & 2\\
    \ATMEGA    &  0 &  22 &  36 &  27 &  9 & \cellcolor{red!40} 107 &  24  & 1\\
    \PIC       & \cellcolor{red!80}225 &  0 &  10 &  18 &  8 &  13 & \cellcolor{red!40} 113  & 1\\
    \MSP       & \cellcolor{red!80}408 &  0 &  34 &  26 &  15 &  13 &  13 & 1\\
    \bottomrule
\end{tabular}
\caption{Model parameters for the different platforms. The letters in brackets show which parameters correspond to the features seen in Figure~\ref{fig:alignment_results}.}
\label{tab:model_parameters}
\end{table*}

\begin{figure}[t]
    \centering
    \includegraphics[width=0.7\linewidth]{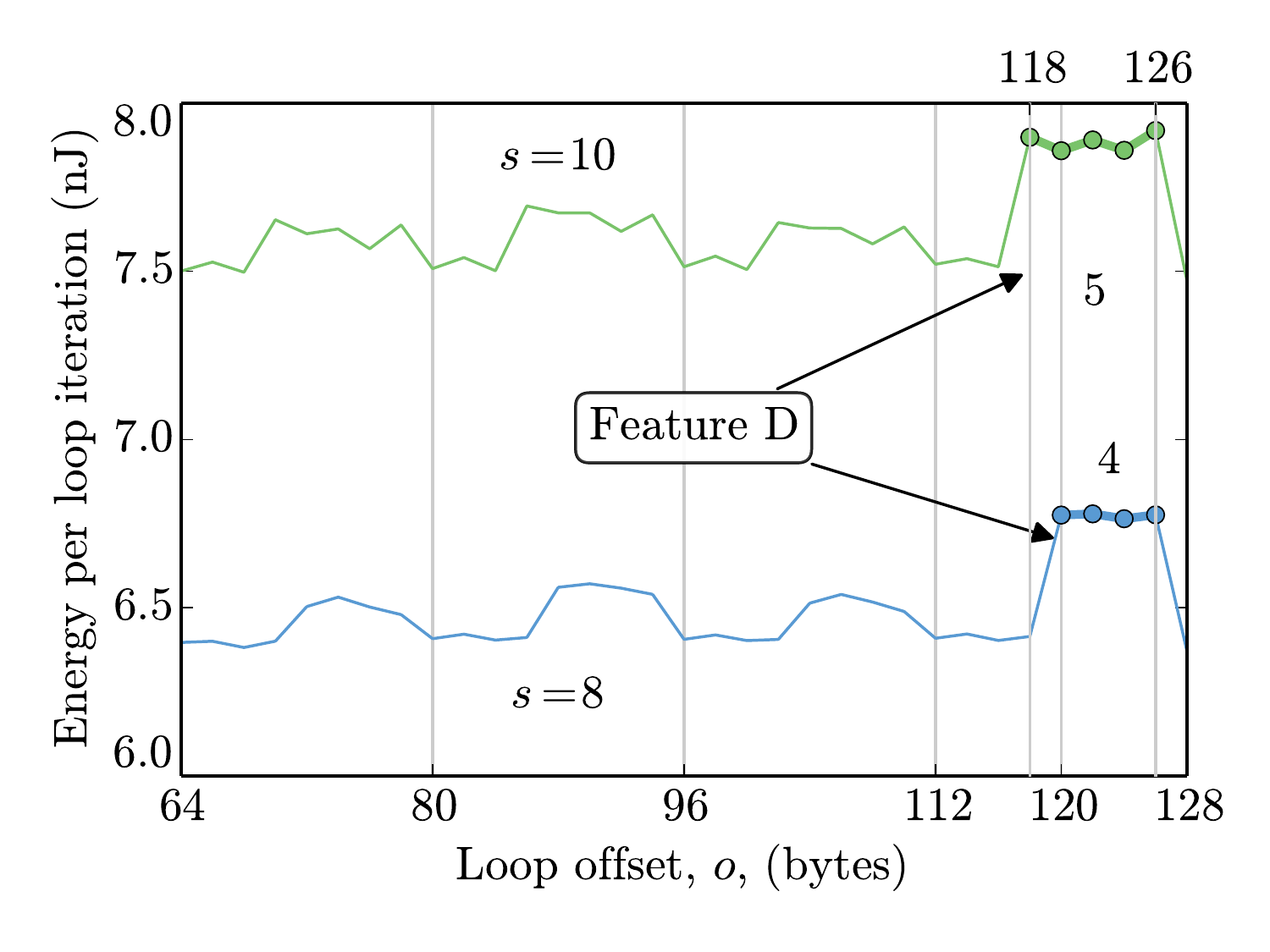}
    \caption{A close up of feature \textbf{D}. This feature highlights the number of offsets which have raised energy consumption is greater than expected.}
    \label{fig:feature_d}
\end{figure}

\begin{description}
    \item[A]
    On the \STMZERO and \STMTHREE platforms the alignment to a 4-byte boundary has a large effect on the energy consumption. This effect occurs because there are 32 bit-lines in the flash for these particular devices, modeled with the $E_2$ parameter. When the loop size is not a multiple of 4 bytes, changing the offset has a small effect --- the same number of 4-byte regions are powered up. The same increase in energy is seen in \PIC and \MSP, however, it is reversed: i.e. only seen when $s$ is not a multiple of 4. This is due to differences in the amount of instruction fetching performed. This was discussed in more detail above, in Section~\ref{sec:fetching}.

    \item[B]
    Increases in energy consumption are seen when the loop straddles multiple 16-byte blocks. This is seen on the \ATMEGA, and on the \MSP to a lesser extent. This is also an artifact of the flash structure: the page is divided into groups of word-lines, totaling 16 bytes. When the loop straddles multiple blocks, additional energy is required to activate all blocks.

    These effects can also be captured by the model, assigning an energy costs to $E_4$ for the 16-byte region.

    \item[C]
    This is the effect of powering up a page in flash, as predicted by the structure of the underlying flash memory and manifests as a spike in energy consumption when the loop spans two flash pages (as shown in Fig.~\ref{fig:loop_overlap}). This can be modeled using Eq.~\ref{eq:flash_line_model} by assigning a large value to the crossing of a 128-byte boundary ($E_7$) --- this causes the changing 128-byte region to have a larger associated energy cost.

    A slightly different pattern is seen for the \PIC and \STMTHREE SoCs. These devices do not have large spikes at 128 bytes, but do at 256 bytes, suggesting that their flash page is 256 bytes long. Consequently, this can be modeled by attributing the energy cost to $E_8$ instead of $E_7$.

    \item[D]
    This feature (also see Fig.~\ref{fig:feature_d}) highlights that the number of tests which have a higher energy consumption is greater than expected for this region. In the highlighted region, there are $k=5$ points with large energy consumption ($10$ bytes, because each instruction is 2 bytes), whereas without instruction fetching, $k=4$. This is a consequence of a loop size $s=10$ having 4 of the alignment tests that would straddle that region boundary. The number of tests expected at a higher energy consumption without instruction fetching is given by:

    \begin{equation}
        k = \frac{s-1}{2}.
    \end{equation}

    However, a larger number of points is seen for all flash-based platforms, due to at least one extra instruction being fetched when a branch is encountered. This results in extra regions being powered up and additional energy consumption, and was discussed previously in Section~\ref{sec:fetching}.
\end{description}

The sixth SoC, \FRAM, sees a completely flat energy profile in the graph. This is due to the use of FRAM instead of flash for this SoC. As discussed in Section~\ref{sec:fram}, the structure of this type of memory different from flash and none of the features seen for the other SoCs appear. As the only differences between this SoC and \MSP are small changes in peripherals and clocking, the characteristics seen in the graph are caused directly by the flash, rather than the processor or SoC interconnect.

\section{Regression}
\label{sec:regression}

\begin{figure}
    \includegraphics[width=\linewidth]{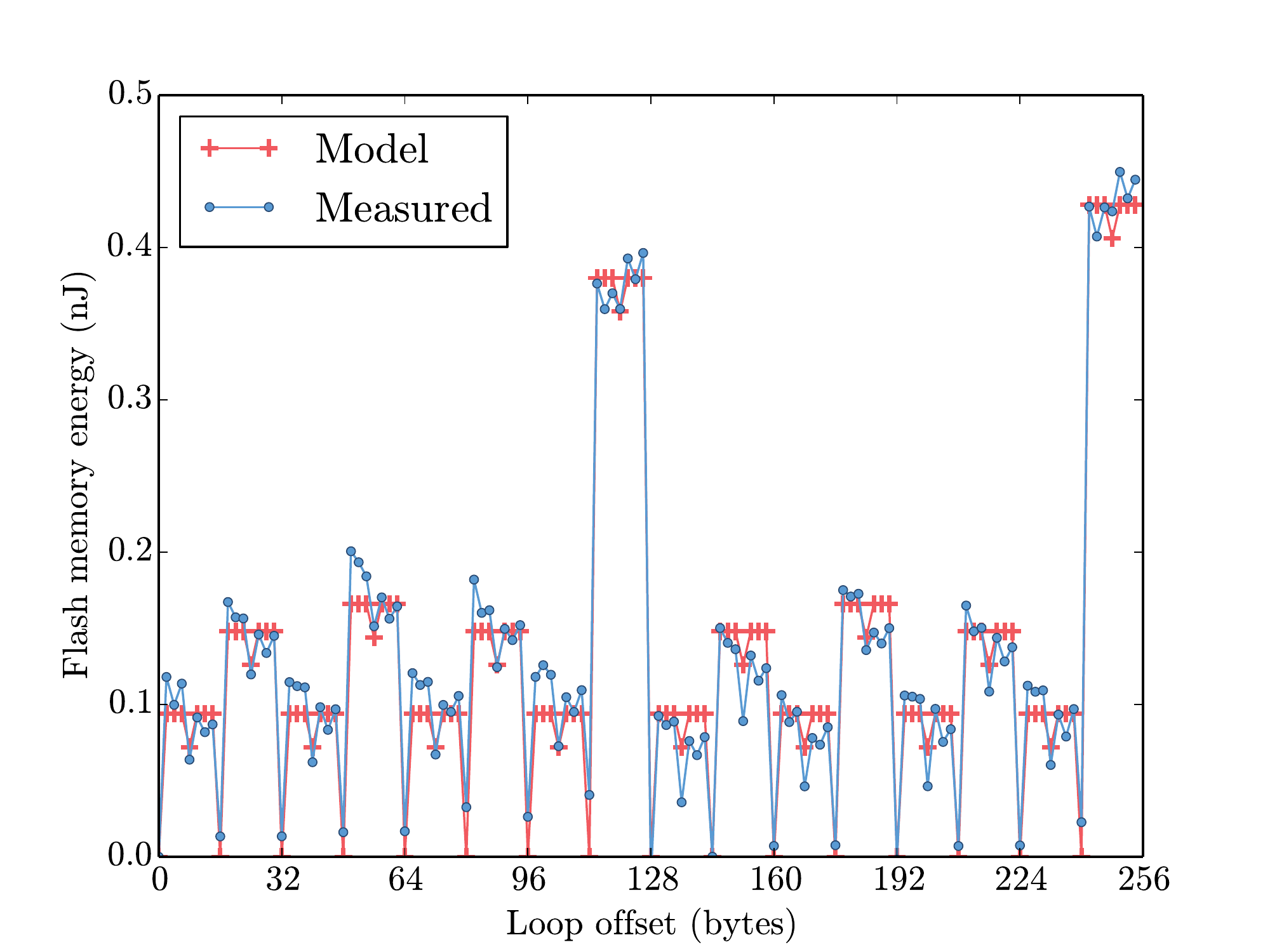}
    \caption{Comparison of model against actual figures for \ATMEGA, $S_{loop}=\{14\}$.}
    \label{fig:model_vs_actual}
\end{figure}

The model (Eq.~\ref{eq:fetch_model}) is fitted to each platform, allowing the energy required to activate different regions to be determined. To find the parameters, linear regression is performed using $O_{loop}=\{0,2,...,256\}$, and $S_{loop}=\{8,10,12,14,16\}$. This allowed the majority of parameters to be fitted (for a total of 645 tests per SoC).

The derived model parameters are shown in Table~\ref{tab:model_parameters}. The highlighted cells are relatively high costs, effects in line with those that appear on the previous graph (Fig.~\ref{fig:alignment_results}) labeled \textbf{A}--\textbf{C}.

An example of the model fitting the previous results is shown in Fig.~\ref{fig:model_vs_actual}, for the \ATMEGA.

The parameters show that alignment is a important issue when executing code --- alignment to a 4-byte boundary will have a large effect on energy consumption if the code is executed frequently. This suggests that for these platforms, there are 32 bit-lines. Also, for many of the SoCs there is a large jump in energy consumption for code which crosses a 128 or a 256-byte region. This is likely due to the size of the pages in flash.

For some SoCs the parameters $E_3$, $E_4$ and $E_5$ have values. This indicates that the flash page may be divided into blocks, with additional energy required to power up the support circuitry in each block.

\subsection*{Model Validation}
\label{sec:validation}

\begin{table}
    \centering
    \begin{tabular}{c r r}
        \toprule
        SoC & \multicolumn{2}{c}{NRMSD \% } \\
        & Cross & Overall \\
        \midrule
        \STMZERO & 27.1 & 22.5 \\
        \STMTHREE & 17.5 & 14.8 \\
        \ATMEGA      & 5.6  & 5.0 \\
        \PIC      & 8.4  & 8.1 \\
        \MSP    & 17.7 & 14.6 \\
        \midrule
        Mean & 13.2 & 11.5 \\
        \bottomrule
    \end{tabular}
    \caption{Normalized root mean square deviation (NRMSD) calculated during validation.}
    \label{tab:validation}
\end{table}

%

\begin{table}
    \centering
    \begin{tabular}{c c c c c c}
        \toprule
        & \multicolumn{2}{c}{NRMSD \%} & \multicolumn{3}{c}{Features}\\
        Program & \STMZERO & \ATMEGA & $B_u$ & $B_c$ & $BB$\\
        \midrule
        \romann{I}   & 14.1 & 8.8  & 1 & 2 & 4 \\  
        \romann{II}\makebox[0pt]{$\hspace{4pt}^\dagger$}  & 19.3 & 6.1  & 1 & 2 & 3 \\  
        \romann{III}\makebox[0pt]{$\hspace{4pt}^\dagger$}& 18.8 & 9.1  & 1 & 2 & 3 \\  
        \romann{IV}  & 14.3 & 5.7  & 2 & 3 & 5 \\  
        \romann{V}\makebox[0pt]{$\hspace{4pt}^\ddagger$}   & 21.7 & 7.4  & 1 & 1 & 2 \\  
        \romann{VI}\makebox[0pt]{$\hspace{4pt}^\ddagger$}  &  9.5 & 7.6  & 1 & 1 & 2 \\  
        \midrule
        Mean  & 15.7 & 7.3 & - & - & - \\
        \bottomrule
    \end{tabular}
    \caption{Normalized root mean square deviation (NRMSD) calculated during validation of complex loops. The table also displays various features of the loops: unconditional branches ($B_u$), conditional ($B_c$), and basic blocks ($BB$).}
    \raggedleft\scriptsize$\dagger\ddagger$ These pairs of tests have the same structure, but different sized blocks.\hspace{1cm}\\

    \label{tab:complex_loop}
\end{table}

The model was validated with cross validation and by testing the model on unseen, more complex loops. The cross validation used $S_{loop}=\{8,10,12,14,16\}$, training on four of the datasets and testing on the remaining. This was repeated for all combinations of datasets, and the average error between observed and predicted data for all SoCs is shown in Table~\ref{tab:validation}. The overall error is also given --- this error indicates how well the model fits the data.

The model performs well for the \ATMEGA and \PIC based SoCs (5.0\% and 8.1\% respectively) and the error is acceptable with the \STMTHREE and \MSP SoCs (14.8\% and 14.6\% respectively). A larger error is seen on the \STMZERO (22.5\%), likely due to the non-trivial instruction fetching and buffering performed. The Cortex-M0 in this SoC has three 32-bit buffers which hold prefetched instructions. The conditions for replenishing these buffers are complex, and dependent on the branching in the instruction stream. If the exact conditions under which these buffers operated is known, the error should be reduced greatly.

The same instruction fetching and buffering is also used for the Cortex-M3 in the \STMTHREE, however it is suspected the branch speculation present in this processor largely cancels out the error, reducing the extraneous memory accesses.

\begin{figure}
    \centering
    \includegraphics[width=\linewidth]{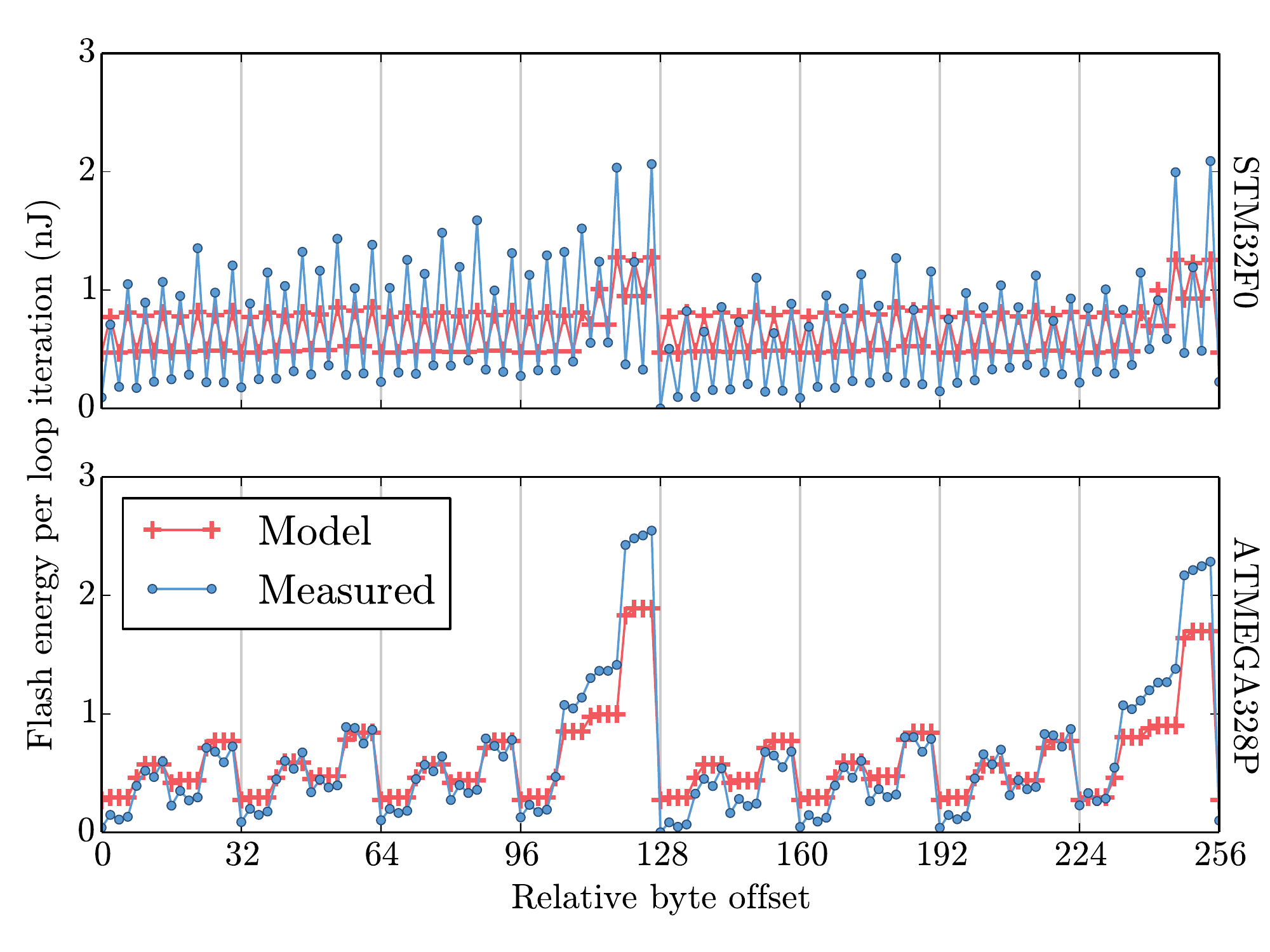}
    \caption{Model against actual data with the \STMZERO and \ATMEGA SoCs, for complex loop \romann{IV}.}
    \label{fig:validation}
\end{figure}

The model was also validated by repeating the loop alignment tests with a set of complex loops, $\romann{I},...,\romann{VI}$. These loops were constructed from example loops seen in BEEBS~\cite{Pallister2013b}, and contain various conditional structures with different numbers of conditional and unconditional branches, number of basic blocks and size of those basic blocks. Using loops generated in this way meant that a good spread of different instruction-level features could be used in just a few loops. The features can be seen in Table~\ref{tab:complex_loop}.

Each of these loops was moved to different locations in memory, and the change in energy this produces was predicted from the model, by applying the model to an instruction trace. Each test had its energy measured on real hardware. The error between observed and predicted data for each loop is shown in Table~\ref{tab:complex_loop}. The platforms containing the \ATMEGA and the \STMZERO processors were chosen because they had the best and worst errors respectively, in the cross validation. Figure~\ref{fig:validation} shows the individual predictions plotted against the measured results for complex loop \romann{IV}.

These results show low error rates for the \ATMEGA processor, indicating that the model predicts the energy consumption of flash memory well, even with more complex loops. The error is higher for the \STMZERO. This is due to the buffering making the sequence of memory accesses very difficult to capture, even with an instruction trace. However, the graph of offset against energy consumption is still qualitatively similar for this processor, meaning that alignment optimizations based on this model should still be effective.

\section{Analysis of Optimization Scope}
\label{sec:optimization_scope}

\begin{figure*}
    \centering
    \begin{tabular}{ c | r r r | r r r | r r r | r r r | r r r}
        \toprule
        \multirow{2}{0.7cm}{\centering Opt. level}& \multicolumn{3}{c|}{\STMZERO} &\multicolumn{3}{c|}{\STMTHREE} & \multicolumn{3}{c|}{\ATMEGA} & \multicolumn{3}{c|}{\PIC} & \multicolumn{3}{c}{\MSP} \\
            & S & W & L & S & W & L & S & W & L & S & W & L & S & W & L\\[0.1em]
        \specialrule{0.3pt}{0pt}{0pt}
        &&&&&&&&&&&&\\[-0.8em]
        \centering \texttt{O0}        & 100& 4.5& 35  &  112& 2.7& 30  &  184& 9.3& 41  &  190& 4.0& 44  & 184\makebox[0pt]{$\hspace{4pt}^\dagger$} & 4.3 & 45 \\
        \centering \texttt{O1}        &  60& 2.6& 26  &   66& 2.3& 17  &  118& 5.8& 37  &   72& 2.0& 30  &   72& 2.0& 30 \\
        \centering \texttt{O2}        & 172& 3.2& 30  &  150& 3.0& 34  &  124& 3.6& 33  &   88& 2.8& 34  &   88& 2.8& 34 \\
        \centering \texttt{O3}        & 528& 2.4& 37  &  370& 1.7& 33  &  690& 1.7& 25  &  882& 2.7& 43  &  882& 2.7& 43 \\
        \centering \texttt{Os}        &  60& 3.4& 26  &   64& 3.6& 24  &   82& 4.0& 35  &   66& 3.0& 34  &   66& 3.0& 34 \\
        \bottomrule
    \end{tabular}
    \caption{The \textit{S} column gives the average loop size in bytes, \textit{W} gives the average percentage increase in code size if each eligible loop was aligned, and \textit{L} gives the percentage of loops that can be realigned to reduce energy consumption for each optimization level.}
    \vspace{1mm}
    \raggedleft\scriptsize$\dagger$ Excluding the rijndael benchmark which failed to compile.\hspace{1cm}
    \label{tab:static_loop}
\end{figure*}

The model derived previously can be used to predict how the alignment of loops and instructions in flash memory can be changed to reduce energy consumption. In this section a benchmark suite is analyzed, and the ability to optimize for each platform is examined.

A possible optimization is to ensure that loops are aligned in a way that minimizes energy. For some of the platforms the greatest model parameter is for the 4-byte region ($E_2$). This can be reduced by ensuring loops are aligned to a 4-byte boundary. This optimization is often seen in modern compilers for performance reasons --- 4 bytes is the bus width of many processors and unaligned accesses often have a performance penalty or are not supported. Alignments at higher boundaries have not been considered, as there is often less or no performance (execution time) benefit.

An energy saving transformation by aligning loops should consider the following items:

\begin{itemize}
    \item \textbf{Estimated minimum number of iterations of the loop.} A trade-off must be made between the cost and the benefit of aligning the loop. This trade-off will be affected by the number of iterations for which the loop is executed.
    \item \textbf{Size of the loop.} The transformation should consider the size of the loop, because large loops will have a lower relative decrease in energy consumption, compared to smaller loops.
    \item \textbf{Space wasted to align the loop.} When aligning the loop to a $k$ byte region, up to $k-1$ bytes may be wasted. The wasted space must be balanced against the benefit of aligning the loop, since blindly aligning every loop to a large boundary could cause a significant increase in code size. It is possible to minimize this by moving infrequently executed basic blocks into the space before the loop.
    \item \textbf{Loop entry distance.} The performance and energy costs of branching into the loop must be weighed against the cost of padding the offset with \texttt{nop}s.
\end{itemize}

Overall the parameters controlling the optimization need to be tuned for each SoC.

The proposed loop alignment optimization was analyzed for its energy saving potential in realistic scenarios. This is performed by analyzing and running the BEEBS~\cite{Pallister2013b} benchmarks, which are designed to expose energy consumption characteristics. These benchmarks were compiled with the latest version of GCC available for each platform.

A tool was written to analyze the binaries resulting from the compilation. This tool uses the algorithm given in \cite{Wei2007} to detect the loops in the program and extracts information about their alignment and size.

The information from the analysis is used to construct an average loop size, percentage increase in code size if all loops were aligned, and the percentage of all loops in the program that could be aligned. Results are shown in Table~\ref{tab:static_loop} for each platform and overall optimization level available in GCC.

In deeply embedded systems, \texttt{O2} and \texttt{Os} are the optimization levels likely to be used. This is because \texttt{O3} can greatly increase the size of the application through function inlining and loop unrolling, and the lower optimizations levels \texttt{O0} and \texttt{O1} often do not provide the required level of performance. For \texttt{O2} and \texttt{Os}, 24--35\% of all loops can be realigned to reduce energy consumption, with only a 2.8--4.0\% increase in code size. It is also expected that there should be minimal increase in execution time (due to the small amount of extra code, outside of the loops).

Overall when applied to code, the analysis shows that this optimization has the potential to reduce energy consumption significantly without greatly increasing code size or execution time.

\section{Related Work}
\label{sec:related_work}

The modeling of energy consumption has been attempted for both embedded systems and larger, more complex processors. Tiwari et al.~\cite{Tiwari1996} constructed an instruction level energy model, assigning an energy cost to each instruction and pair of instructions. This model had an extra parameter, to denote `other' effects --- anything not directly related to an instruction's execution. This would include effects as seen in this paper, as well as caching and I/O. A further study~\cite{Steinke2001} created a more detailed model, including terms for the memory energy. However, the terms only considered the hamming weight of the address, and the hamming distance between consecutive addresses. Other studies have looked at these other parts of the systems, including caches~\cite{Chandra2008}, DRAM~\cite{Vogelsang2010} and peripherals~\cite{Celebican2004}.

Flash memory's power consumption has been modeled at a low level~\cite{Mohan2013}. This study constructed a detailed model of flash power consumption derived from the transistor and layout level information. Their model was validated against physical measurements of a flash chip, but requires detailed knowledge of the exact structure of the flash. Additionally, this model is not applicable to embedded flash, which has a different, simpler structure. Joo et al.~\cite{Joo2008} characterize the energy required to write to multi-level cell flash devices, and develop an energy aware compression method to exploit the value dependent nature of the energy consumption.

Software approaches have been considered frequently in optimizing the energy consumption of the memory hierarchy in non-embedded devices. Kim et al.~\cite{Kim2000} model the memory hierarchy and evaluate different cache configurations and algorithms, finding that compilers were successful in reducing the energy consumption of data accesses. However, by doing this the instruction-access energy increased. This effect was also seen in~\cite{Aa2004}. Other studies have attempted to optimize the data structure layout in memory to reduce their impact on the cache~\cite{Chilimbi1999}.

In embedded devices, efficiently using scratchpad memory has been considered in~\cite{Gauthier2010,Chen2003,Verma2004}, finding that significant savings in energy could be achieved. These studies exploit the fact that scratchpad memory is faster to access, due to its proximity to the processor. Various algorithms for deciding which items of code and data should be stored in this memory are given, and shown to save significant amounts of energy and execution time. Other optimizations have focused on reducing the number of memory operations~\cite{Zambreno2002}. Since memory operations are typically more energy intensive than processing operations, reducing memory operations leads to an overall lower energy consumption.

Other software optimizations targeting energy have considered automatically inserting \texttt{idle} instructions~\cite{Seth2001}, instruction scheduling~\cite{Parikh2000}, use of SIMD~\cite{Ibrahim2009} and exploiting differences in functional units~\cite{Chakrapani2001}.

Overall, there has not been much work studying embedded flash memory's effect on code execution, particularly for energy consumption. This has likely been overlooked, as this is no performance gain from aligning code. The techniques presented in this paper represent a first step towards being able to exploit the energy consumption characteristics of embedded flash.


\section{Conclusion}
\label{sec:conclusion}

In this paper we discussed the structure of embedded flash memory and show how the internal structure of the flash can have a significant effect on the energy consumption of the overall system.

In Section~\ref{sec:loop_alignment}, altering the alignment of loops exposed significant changes in the energy consumption --- up to 15\% change in total energy consumption on the \STMZERO. This effect was also seen on other SoCs to a lesser, but still significant degree. A generic model was created to predict this energy consumption due to code positioning in flash. This model considered the circuit state-change overhead between sequential memory accesses, by assigning an energy cost to accessing each $2^k$-byte region.

The parameters for the model were derived for five different SoCs, and these parameters correlated to the structure of the underlying flash. The model was validated with these parameters, using cross validation on the loop alignment tests. For the SoCs with the largest (\STMZERO) and smallest (\ATMEGA) errors, more extensive validation was performed, using loops with complex control structures and conditional branching. The error for both platforms remained similar to the cross validation, indicating that the model can cope with arbitrary code. While the error for the \STMZERO SoC was large (15.7\%), the observation against the prediction was qualitatively similar, thus allowing a more efficient code placement to be predicted with this model.

The sixth SoC (\FRAM) used FRAM technology instead of flash, but was otherwise similar to \MSP. The code alignment did not have any significant effect on the energy consumption in this processor, as expected from the random access nature of the FRAM.

The potential of optimizing code using based on this model was discussed. The transformation would ensure that the start of loops were aligned to a $2^k$-byte boundary, reducing the number of $k$-byte boundaries crossed by the code. The value of $k$ would be chosen based on the model's parameters for the target SoC. The proposed optimization was shown to be applicable to 20--40\% of loops in a variety of benchmarks and cause less than 4\% increase in code size on average. This provides guidance when programming in assembly code, where the programmer may have direct control over where the code is placed. Additionally this optimization could be implemented in a compiler, to automatically align loops created in high level languages.

Overall there is the possibility to save energy in a previously unconsidered way, exploiting the structure of embedded flash. The given model can predict the energy due to the flash. This enables the design of an optimization to reduce energy consumption. 

\acks

This study was funded by Embecosm and was also partly sponsored by EPSRC's Doctoral Training Account EP/K502996/1 and a HiPEAC internship (to J.P.).

\bibliographystyle{plain}

\bibliography{library}

\begin{thebibliography}{10}

\bibitem{Aa2004}
Tom~Vander Aa, Murali Jayapala, Francisco Barat, Henk Corporaal, Francky
  Catthoor, and Geert Deconinck.
\newblock {Instruction and Data Memory Energy Trade-off using a High-level
  Model Steering the Loop Merging Transformations}.
\newblock In {\em Proceedings of 2nd Workshop on Optimizations for DSP and
  Embedded Systems}, pages 1--7, 2004.

\bibitem{Abari2008}
R.~Abari, S.~Basu, T.~Chen, A.~Chatterjee, S.~Farshchi, T.~G. Croda, R.~J.
  Herrick, B.~M. Hammerli, M.~S. Newman, and S.~V. Kartalopoulos.
\newblock {\em {Nonvolatile Memory Technologies with Emphasis on Flash}}.
\newblock IEEE Press Series on Microelectronic Systems. John Wiley \& Sons,
  Inc., Hoboken, NJ, USA, 1st edition, December 2008.

\bibitem{Calder1998}
Brad Calder, Chandra Krintz, Simmi John, and Todd Austin.
\newblock {Cache-conscious data placement}.
\newblock {\em ACM SIGPLAN Notices}, 33(11):139--149, November 1998.

\bibitem{Cavaleri2004}
Paola Cavaleri, Bruno Leconte, S\'{e}bastien Zink, and Jean Devin.
\newblock {Page-erasable flash memory}, 2004.

\bibitem{Celebican2004}
Ozgur Celebican, Tajana~Simunic Rosing, and Vincent~J. {Mooney III}.
\newblock {Energy Estimation of Peripheral Devices in Embedded Systems}.
\newblock In {\em Proceedings of the 14th ACM Great Lakes symposium on VLSI},
  pages 430--435. ACM, 2004.

\bibitem{Chakrapani2001}
L.~N. Chakrapani, P.~Korkmaz, V.~J. {Mooney III}, K.~V. Palem, K.~Puttaswamy,
  and W.~F. Wong.
\newblock {The emerging power crisis in embedded processors: what can a (poor)
  compiler do?}
\newblock In {\em Proceedings of the 2001 international conference on
  Compilers, Architecture, and Synthesis for Embedded Systems}. ACM, 2001.

\bibitem{Chandra2008}
Lokesh Chandra and Sourav Roy.
\newblock {Estimation of energy consumed by software in processor caches}.
\newblock In {\em 2008 IEEE International Symposium on VLSI Design, Automation
  and Test (VLSI-DAT)}, pages 21--24. IEEE, April 2008.

\bibitem{Chen2003}
G.~Chen, I.~Kadayif, W.~Zhang, M.~Kandemir, and I.~Kolcu.
\newblock {Compiler-Directed Management of Instruction Accesses}.
\newblock In {\em Proceedings Euromicro Symposium on Digital System Design}.
  IEEE Computer Society, 2003.

\bibitem{Chilimbi1999}
Trishul~M. Chilimbi, James~R. Larus, and Mark~D. Hill.
\newblock {Cache-conscious structure layout}.
\newblock {\em ACM SIGPLAN Notices}, pages 1--12, 1999.

\bibitem{Gauthier2010}
Lovic Gauthier, Tohru Ishihara, Hideki Takase, Hiroyuki Tomiyama, and Hiroaki
  Takada.
\newblock {Minimizing Inter-Task Interferences in Scratch-Pad Memory Usage for
  Reducing the Energy Consumption of Multi-Task Systems}.
\newblock In {\em Proceedings of the 2010 international conference on
  Compilers, Architectures and Synthesis for Embedded Systems}. ACM, 2010.

\bibitem{Ibrahim2009}
Mostafa E.~A. Ibrahim, Markus Rupp, and Hossam A.~H. Fahmy.
\newblock {Code transformations and SIMD impact on embedded software
  energy/power consumption}.
\newblock In {\em 2009 International Conference on Computer Engineering \&
  Systems}, pages 27--32. IEEE, December 2009.

\bibitem{Joo2008}
Yongsoo Joo, Youngjin Cho, Donghwa Shin, Jaehyun Park, and Naehyuck Chang.
\newblock {An energy characterization platform for memory devices and
  energy-aware data compression for multilevel-cell flash memory}.
\newblock {\em ACM Transactions on Design Automation of Electronic Systems},
  13(3):1--29, July 2008.

\bibitem{Kandemir2002a}
Mahmut Kandemir, N.~Vijaykrishnan, and Mary~Jane Irwin.
\newblock {Compiler optimizations for low power systems}.
\newblock In {\em Power Aware Computing}, pages 191----210. 2002.

\bibitem{Kato2010}
Yoshihisa Kato, Hiroyuki Tanaka, Kazunori Isogail, Kazuhiro Kaibaral, Yukihiro
  Kaneko, Yasuhiro Shimadal, Matt Brubaker, Jolanta Celinska, Larry~D.
  Mcmillan, and Carlos~A. {Paz De Araujo}.
\newblock {Embedded FeRAM Challenges in the 65-nm Technology Node and Beyond}.
\newblock In {\em International Symposium on Applications of Ferroelectrics},
  pages 1--4. IEEE, 2010.

\bibitem{Kim2000}
H.~S. Kim, Mary~Jane Irwin, N.~Vijaykrishnan, and M.~Kandemir.
\newblock {Effect of compiler optimizations on memory energy}.
\newblock In {\em IEEE Workshop on Signal Processing Systems}, pages 663--672.
  IEEE, 2000.

\bibitem{Mohan2013}
Vidyabhushan Mohan, Trevor Bunker, Laura~M. Grupp, Sudhanva Gurumurthi,
  Mircea~R. Stan, and Steven Swanson.
\newblock {Modeling Power Consumption of NAND Flash Memories Using FlashPower}.
\newblock {\em IEEE Transactions on Computer-Aided Design of Integrated
  Circuits and Systems}, 32(7):1031--1044, 2013.

\bibitem{Pallister2013b}
James Pallister, Simon Hollis, and Jeremy Bennett.
\newblock {BEEBS: Open Benchmarks for Energy Measurements on Embedded
  Platforms}.
\newblock 2013.

\bibitem{Parikh2000}
A.~Parikh, Mahmut Kandemir, N.~Vijaykrishnan, and Mary~Jane Irwin.
\newblock {Instruction scheduling based on energy and performance constraints}.
\newblock In {\em Proceedings. IEEE Computer Society Workshop on VLSI}, pages
  37--42. IEEE Comput. Soc, 2000.

\bibitem{RenesasElectronics2014}
{Renesas Electronics}.
\newblock {Renesas Electronics Develops Industry's First 28nm Embedded Flash
  Memory Technology for Microcontrollers}, 2014.

\bibitem{Seth2001}
Anil Seth, R.~B. Keskar, and R.~Venugopal.
\newblock {Algorithms for energy optimization using processor instructions}.
\newblock In {\em CASES '01 Proceedings of the 2001 international conference on
  Compilers, Architecture, and Synthesis for Embedded Systems}, page 195, New
  York, New York, USA, 2001. ACM.

\bibitem{Steinke2001}
Stefan Steinke, Markus Knauer, Lars Wehmeyer, and Peter Marwedel.
\newblock {An accurate and fine grain instruction-level energy model supporting
  software optimizations}.
\newblock In {\em Proceedings of PATMOS}, 2001.

\bibitem{STMicroelectronics2005}
STMicroelectronics.
\newblock {High density NAND flash memories}, 2005.

\bibitem{Tiwari1996}
Vivek Tiwari, Sharad Malik, Andrew Wolfe, and Mike {Tien-Chien Lee}.
\newblock {Instruction level power analysis and optimization of software}.
\newblock {\em Journal of VLSI Signal Processing Systems for Signal, Image, and
  Video Technology}, 13(2-3):223--238, 1996.

\bibitem{Verma2004}
Manish Verma, Lars Wehmeyer, and Peter Marwedel.
\newblock {Dynamic Overlay of Scratchpad Memory for Energy Minimization}.
\newblock In {\em International Conference on Hardware/software Codesign and
  System Synthesis}. ACM, 2004.

\bibitem{Vogelsang2010}
Thomas Vogelsang.
\newblock {Understanding the Energy Consumption of Dynamic Random Access
  Memories}.
\newblock In {\em 2010 43rd Annual IEEE/ACM International Symposium on
  Microarchitecture}, pages 363--374. IEEE, December 2010.

\bibitem{Wei2007}
Tao Wei, Jian Mao, Wei Zou, and Yu~Chen.
\newblock {A new algorithm for identifying loops in decompilation}.
\newblock {\em Static Analysis}, pages 170--183, 2007.

\bibitem{Zambreno2002}
Joseph Zambreno, Mahmut Kandemir, and Alok Choudhary.
\newblock {Enhancing compiler techniques for memory energy optimizations}.
\newblock {\em Embedded Software}, pages 364--381, 2002.

\end{thebibliography}

\end{document}